\documentclass[12pt]{article}

\usepackage[dvips]{graphics}

\def\beq   {\begin{equation}}
\def\eeq   {\end{equation}}
\def\beqd  {\begin{displaymath}}
\def\eeqd  {\end{displaymath}}
\def\beqaa {\begin{eqnarray}}
\def\eeqaa {\end{eqnarray}}

\def\noi {\noindent}

\def\ti  {\tilde}

\def\sq  {\ti q}
\def\st  {\ti t}
\def\sb  {\ti b}
\def\sg  {\ti g}
\def\sa  {\ti \tau}
\def\sn  {\ti \nu}

\def\nt  {\tilde\chi^0}
\def\ch  {\tilde\chi^\pm}
\def\chp {\tilde\chi^+}
\def\chm {\tilde\chi^-}

\def\a   {\alpha}
\def\b   {\beta}
\def\t   {\theta}
\def\tst {\theta_{\st}}
\def\tsb {\theta_{\sb}}

\def\sz{\ifmmode{\tilde{\chi}^0} \else{$\tilde{\chi}^0$} \fi}
\def\sw{\ifmmode{\tilde{\chi}} \else{$\tilde{\chi}$} \fi}

\newcommand{\gsim}{\;\raisebox{-0.9ex}
           {$\textstyle\stackrel{\textstyle >}{\sim}$}\;}

\begin{document}
\pagestyle{empty}

\vspace*{-1cm} 
\begin{flushright}
  TGU-26 \\
  UWThPh-2000-46 \\
  hep-ph/0012021
\end{flushright}

\vspace*{1.4cm}

\begin{center}

{\Large {\bf
Impact of bosonic decays on the search\\
for \boldmath{$\tilde t_1$} and \boldmath{$\tilde b_1$} squarks
}}\\

\vspace{10mm}

{\large 
K.~Hidaka$^a$ and A.~Bartl$^b$}



\vspace{6mm}

\begin{tabular}{l}
$^a${\it Department of Physics, Tokyo Gakugei University, Koganei,
Tokyo 184--8501, Japan}\\
$^b${\it Institut f\"ur Theoretische Physik, Universit\"at Wien, A-1090
Vienna, Austria}
\end{tabular}



\end{center}

\vfill

\begin{abstract} 
We perform a detailed study of the decays of the lighter top and bottom
squarks ($\st_1$ and $\sb_1$)
in the Minimal Supersymmetric Standard Model (MSSM). 
We show that the decays into Higgs or gauge bosons, i.e. 
$\st_1 \to \sb_1 + (H^+ \ \mbox{or} \ W^+)$ and
$\sb_1 \to \st_1 + (H^- \ \mbox{or} \ W^-)$, 
can be dominant in a wide range of the MSSM parameters
due to the large Yukawa couplings and mixings of $\st$ and $\sb$. 
Compared to the decays into fermions, such as 
$\st_1 \to t + (\nt_i \ \mbox{or} \ \sg)$ and $\st_1 \to b + \chp_j$, 
these bosonic decay modes can have significantly different decay 
distributions. We also show that the effect of the supersymmetric 
QCD running of the quark and squark parameters 
on the $\st_1$ and $\sb_1$ decay branching ratios is quite dramatic.
These could have an important impact on the search for $\st_1$ 
and $\sb_1$ and the determination of the MSSM parameters 
at future colliders.
\end{abstract}

\newpage
\pagestyle{plain}
\setcounter{page}{2}


In the Minimal Supersymmetric Standard Model (MSSM) \cite{Haber-Kane} 
supersymmetric (SUSY) partners of all the Standard Model (SM) particles  
are introduced. In order to solve the hierarchy, fine-tuning and 
naturalness problems, the SUSY particle masses have to be less 
than O(1 TeV). Hence the discovery of all SUSY partners and the 
study of their properties are essential for testing the MSSM.  
Future colliders, such as the Large Hadron Collider (LHC), the upgraded 
Tevatron, $e^+e^-$ linear colliders, and $\mu^+\mu^-$ colliders will 
push the discovery reach for the SUSY particles up to the TeV mass range 
and allow for precise measurement of the MSSM parameters.

\noi
In this article we perform a phenomenological study concerning the 
search for the SUSY partners of the top (t) and bottom (b) quarks 
(i.e. stops ($\st$) and sbottoms ($\sb$)). These particles may have 
properties which are very different from those of the squarks of 
the other two generations due to the large top and bottom Yukawa 
couplings. Production and decays of $\st$ and $\sb$ were 
studied in \cite{BMP, Porod-PHD, Sopczak, stop2decay}. 
Stops (sbottoms) have two mass 
eigenstates $\st_{1,2}$ ($\sb_{1,2}$) with $m_{\st_1}<m_{\st_2}$ 
($m_{\sb_1}<m_{\sb_2}$). Here we focus on $\st_1$ and $\sb_1$.
Like other squarks, they can decay into fermions, i.~e. 
a quark plus a gluino ($\sg$), neutralino ($\nt_i$) or chargino ($\ch_j$): 
\beq
  \begin{array}{lcl}
  \st_1 \to t\,+\,(\sg \ \mbox{or} \ \nt_i)\,, &\hspace{3mm}& 
  \sb_1 \to b\,+\,(\sg \ \mbox{or} \ \nt_i)\,, \\
  \st_1 \to b\,+\,\chp_j\,, &\hspace{3mm}& 
  \sb_1 \to t\,+\,\chm_j\,,
  \end{array}
  \label{eq:fmodes}
\eeq
with $i=1,...,4$ and $j=1,2$. In addition, they can also decay into
bosons \cite{BMP, Porod-PHD}, i.~e. a squark plus a Higgs or gauge boson:
\beq
  \begin{array}{lcl}
    \st_1 \to \sb_1 + (H^+ \ \mbox{or} \ W^+)\,, &\hspace{3mm}&
    \sb_1 \to \st_1 + (H^- \ \mbox{or} \ W^-)\,.
  \end{array}
  \label{eq:bmodes}
\eeq
The decays of Eq.~(\ref{eq:bmodes}) are possible in case the mass 
difference between $\st_1$ and $\sb_1$ is sufficiently large. 

\noi
In the present article we extend the analysis of \cite{BMP, Porod-PHD}. 
We point out that the bosonic decays of $\st_1$ and $\sb_1$ of 
Eq.~(\ref{eq:bmodes}) can be dominant in a large region of 
the MSSM parameter space due to large top and bottom Yukawa couplings 
and large $\st$ and $\sb$ mixing parameters.  
This dominance of the bosonic decays over the conventional 
fermionic decays of Eq.~(\ref{eq:fmodes}) could have an important impact 
on searches for $\st_1$ and $\sb_1$ at future colliders.
An analogous study for $\st_2$, $\sb_2$, $\sa_2$, and $\sn_\tau$ was 
performed in \cite{stop2decay, stau2decay}, where $\sa_2$ and $\sn_\tau$ 
are the sleptons of the third generation. In addition we show that the effect 
of SUSY-QCD running of the quark and squark parameters \cite{improvedQCDcorr} 
on the decay branching ratios of $\st_1$ and $\sb_1$ is 
quite dramatic.


First we summarize the MSSM parameters in our analysis. 
In the MSSM the squark sector is specified by the mass matrix in 
the basis $(\sq_L^{},\sq_R^{})$ with $\sq=\st$ or 
$\sb$~\cite{Ellis-Rudaz, Gunion-Haber}
\begin{equation}
  {\cal M}^2_{\sq}= 
     \left( 
            \begin{array}{cc} 
                m_{\sq_L}^2 & a_q m_q \\
                a_q m_q     & m_{\sq_R}^2
            \end{array} 
     \right)       
  \label{eq:f}
\end{equation}
with
\begin{eqnarray}
  m_{\sq_L}^2 &=& M_{\ti Q}^2 
                  + m_Z^2\cos 2\beta\,(I_3^{q_L} - e_q\sin^2\t_W) 
                  + m_q^2, \label{eq:g} \\
  m_{\sq_R}^2 &=& M_{\{\ti U,\ti D\}}^2  
                  + m_Z^2 \cos 2\b\, e_q\, \sin^2\t_W + m_q^2, 
                                              \label{eq:h}\\[2mm]
  a_q m_q     &=& \left\{ \begin{array}{l}
                     (A_t - \mu\cot\beta)\;m_t~~(\sq=\st)\\
                     (A_b - \mu\tan\beta)\,m_b~~(\sq=\sb) \, .
                          \end{array} \right. \label{eq:i}
\end{eqnarray}
Here $I_3^q$ is the third component of the weak isospin and $e_q$ the 
electric charge of the quark $q$.
$M_{\ti Q,\ti U,\ti D}$ and $A_{t,b}$ are soft SUSY--breaking 
parameters, $\mu$ is the higgsino mass parameter, 
and $\tan\b = v_2/v_1$ with $v_1$ $(v_2)$ being the vacuum 
expectation value of the Higgs field $H_1^0$ $(H_2^0)$. 
%
%
We treat $M_{\ti Q,\ti U,\ti D}$ and $A_{t,b}$ as free parameters 
since the ratios $M_{\ti U}/M_{\ti Q}$, $M_{\ti D}/M_{\ti Q}$ and 
$A_t/A_b$ are highly model-dependent.
By diagonalizing the matrix (\ref{eq:f}) one gets the mass eigenstates 
$\sq_1^{}=\sq_L^{}\cos\t_{\sq}+\sq_R^{}\sin\t_{\sq}$, 
$\sq_2^{}=-\sq_L^{}\sin\t_{\sq}+\sq_R^{}\cos\t_{\sq}$ 
with the mass eigenvalues $m_{\sq_1}$, $m_{\sq_2}$ ($m_{\sq_1}<m_{\sq_2}$) 
and the mixing angle $\t_{\sq}$ $(-\frac{\pi}{2} < \t_{\sq} \leq 
\frac{\pi}{2})$.
As can be seen, sizable mixing effects can be expected in the stop sector 
due to the large top quark mass. Likewise, 
$\sb_L$--$\sb_R$ mixing may be important for large $\tan\beta$. \\
The properties of the charginos $\ch_i$ ($i=1,2$; $m_{\ch_1}<m_{\ch_2}$) 
and neutralinos $\nt_k$ ($k=1,...,4$; $m_{\nt_1}< ...< m_{\nt_4}$)  
are determined by the parameters $M$, $M'$, $\mu$ and $\tan\b$, where $M$ 
and $M'$ are the SU(2) and U(1) gaugino mass parameters, respectively. 
Assuming gaugino mass unification we take $M'=(5/3)\tan^2\t_W M$ and
$m_{\sg}=(\alpha_s(m_{\sg})/\alpha_2)M$ with $m_{\sg}$ being the gluino 
mass. 
The masses and couplings of the Higgs bosons $h^0, ~H^0, ~A^0$ and $H^{\pm}$, 
including leading radiative corrections, are fixed by 
$m_A,~\tan\beta,~\mu,~m_t,~m_b,~M_{\ti Q},~M_{\ti U},~M_{\ti D}, 
~A_t,$ and $A_b$. 
$H^0$ ($h^0$) and $A^0$ are the heavier (lighter) CP--even and CP--odd 
neutral Higgs bosons, respectively. 
For the radiative corrections to the $h^0$ and $H^0$ masses 
and their mixing angle $\alpha$ we use the formulae of 
Ref.~\cite{Ellis-Ridolfi-Zwirner}; for those to  $m_{H^+}$ we follow 
Ref.~\cite{Brignole} 
\footnote{Notice that \cite{Ellis-Ridolfi-Zwirner, Brignole} have a sign 
convention for the parameter $\mu$ opposite to the one used here.}.

The widths of the squark decays into Higgs and gauge bosons 
$(H^\pm \ \mbox{and} \ W^\pm)$ are given by \cite{BMP}:
\beqaa
  \Gamma (\sq_1^{} \to \sq_1^\prime \, H^\pm) &=& 
    \frac{\kappa_{H^+}^{}}{16\pi\,m_{\sq_1}^3}\;G_{H^+}^{\ 2} , 
                                             \label{eq:width1} \\
  \Gamma (\sq_1^{} \to \sq_1^\prime \, W^\pm) &=& 
    \frac{\kappa_{W^+}^3}{16\pi\,m_W^2\,m_{\sq_1}^3}\;C_{W^+}^{\ 2}.
                                                 \label{eq:width2}
\eeqaa
Here $(\sq_1,\sq_1^\prime)=(\st_1,\sb_1) \ \mbox{or} \ (\sb_1,\st_1)$. 
$\kappa_{X}^{}\equiv\kappa(m_{\sq_1}^2,m_{\sq_1^\prime}^2,m_X^2)$
is the usual kinematic factor, 
$\kappa(x,y,z)=(x^2+y^2+z^2-2xy-2xz-2yz)^{1/2}$. 
Notice an extra factor $\kappa^2 / m_W^2$ for the gauge boson mode.
The $G_{H^+}$ denotes the squark coupling to $H^+$ 
and $C_{W^+}$ that to $W^+$.
Their complete expressions, as well as the widths 
of the fermionic modes, are given in \cite{BMP, Porod-PHD}. 

From Eqs.(3-6) we see that $m_{\st_{1,2}} \sim M_{\ti Q,\ti U}$ and 
$m_{\sb_{1,2}} \sim M_{\ti Q,\ti D}$ in case $M_{\ti Q,\ti U,\ti D}$ 
are large relatively to the other parameters. 
In this case, for $M_{\ti U} > M_{\ti Q} \gg M_{\ti D}$ 
($M_{\ti D} > M_{\ti Q} \gg M_{\ti U}$) we have $m_{\st_1} \gg m_{\sb_1}$ 
($m_{\sb_1} \gg m_{\st_1}$), which may allow the bosonic decays of 
Eq.(\ref{eq:bmodes}). 
Hence in this article we consider two different patterns of the squark mass 
spectrum: $m_{\st_1} \gg m_{\sb_1}$ with ($\st_1$,$\sb_1$) $\sim$ 
($\st_L$,$\sb_R$) for $M_{\ti U} \gg M_{\ti Q} \gg M_{\ti D}$, and 
$m_{\sb_1} \gg m_{\st_1}$ with ($\st_1$,$\sb_1$) $\sim$ 
($\st_R$,$\sb_L$) for $M_{\ti D} \gg M_{\ti Q} \gg M_{\ti U}$. 
Note here that in the former (latter) pattern the condition 
$M_{\ti U} \gg M_{\ti Q}$ ($M_{\ti D} \gg M_{\ti Q}$) ensures 
$\st_1 \sim \st_L$ ($\sb_1 \sim \sb_L$), which eventually enhances the 
bosonic decays of Eq.(\ref{eq:bmodes}) as we will see below. Thus the bosonic decays 
considered here are basically the decays of $\st_L$ into $\sb_R$ and 
$\sb_L$ into $\st_R$.

The leading terms of the squark couplings to $H^\pm$ and $W^\pm$ are 
given by 
\beqaa
G_{H^+}=G(\st_1 \sb_1 H^\pm) &\sim& 
               h_t(\mu\sin\b + A_t\cos\b)\sin\tst\cos\tsb  \nonumber \\
 & & {} + h_b(\mu\cos\b + A_b\sin\b)\cos\tst\sin\tsb, \label{eq:GH+} \\
C_{W^+}=C(\st_1 \sb_1 W^\pm) &\sim& \frac{g}{\sqrt{2}}\cos\tst\cos\tsb.
                                                         \label{eq:CW+}
\eeqaa
The Yukawa couplings $h_{t,b}$ are given as 
\beqaa
h_t=g m_t/(\sqrt{2} m_W \sin\b), \ \ \ 
h_b=g m_b/(\sqrt{2} m_W \cos\b). \label{eq:h_t,b}
\eeqaa
The Higgs bosons $H^\pm$ couple mainly to $\sq_L \sq_R^\prime$ 
combinations. These couplings are proportional to the Yukawa couplings 
$h_{t,b}$ and the squark mixing parameters $A_{t,b}$ and $\mu$, as can be seen 
in Eq.(\ref{eq:GH+}). Hence the widths of the squark decays into $H^\pm$ 
may be large for large $A_{t,b}$ and $\mu$. Note here that the squark 
mixing angles $\t_{\sq}$ themselves depend on $A_{t,b}$, $\mu$ and $\tan\b$. 
In contrast, the gauge bosons $W^\pm$ couple only to $\sq_L \sq_L^\prime$, 
which results in suppression of the decays into $W^\pm$. However, this 
suppression is largely compensated by the extra factor $\kappa^2 / m_W^2$ 
in Eq.(\ref{eq:width2}) which is very large for $m_{\sq_1}-m_{\sq_1^\prime} 
\gg m_W$. (This factor stems from the contribution of the longitudinally 
polarized gauge boson radiation ($\sq_1 \to \sq_1^\prime W_L^\pm$).) 
Hence the widths of the squark decays into $W^\pm$ may be large for a 
sizable $\sq_L^\prime-\sq_R^\prime$ mixing term $a_{q^\prime} m_{q^\prime}$. 
On the other hand, the fermionic decays are not enhanced for large $A_{t,b}$ 
and $\mu$. Therefore the branching ratios of the bosonic decays of 
Eq.(\ref{eq:bmodes}) are expected to be large for large $A_{t,b}$ and $\mu$ 
if the gluino mode is kinematically forbidden. As for the $\tan\b$ dependence, 
we expect that the branching ratio $B(\st_1 \to \sb_1 + (H^+,W^+))$ increases 
with increasing $\tan\b$ while $B(\sb_1 \to \st_1 + (H^-,W^-))$ is rather 
insensitive to $\tan\b$. The reason for this is as follows:\\
\\
(i)$\st_1$ decay;\\
As $(\st_1,\sb_1) \sim (\st_L,\sb_R)$ in this decay, the coupling 
$G_{H^+}=G(\st_1 \sb_1 H^+) \sim G(\st_L \sb_R H^+) \propto h_b 
\propto \tan\b$. Hence we expect that the width 
$\Gamma(\st_1 \to \sb_1 H^+)$ ( and hence $B(\st_1 \to \sb_1 H^+)$) 
increases with increasing $\tan\b$. The coupling 
$C_{W^+}=C(\st_1 \sb_1 W^+)$ increases with increasing 
$\sb_L - \sb_R$ mixing term $a_b m_b$ which increases with increasing 
$\tan\b$. Hence we expect an increase of $B(\st_1 \to \sb_1 W^+)$ as 
$\tan\b$ increases. \\
(ii)$\sb_1$ decay;\\
As $(\st_1,\sb_1) \sim (\st_R,\sb_L)$ in this decay, the coupling 
$G_{H^-}=G(\st_1 \sb_1 H^-) \sim G(\st_R \sb_L H^-) \propto h_t$. 
Hence we expect that $B(\sb_1 \to \st_1 H^-)$ is rather 
insensitive to $\tan\b$ ( for $\tan\b \gsim 3$). The coupling 
$C_{W^-}=C(\st_1 \sb_1 W^-)$ increases with increasing 
$\st_L - \st_R$ mixing term $a_t m_t$ which is fairly 
insensitive to $\tan\b$ (for $\tan\b \gsim 3$ ). 
Hence we expect a mild dependence of $B(\sb_1 \to \st_1 W^-)$ on 
$\tan\b$.

We now turn to the numerical analysis of the $\st_1$ and $\sb_1$ 
decay branching ratios. We calculate the widths of all possibly 
important two-body decay modes of Eqs. (\ref{eq:fmodes}) and (\ref{eq:bmodes}). 
Three-body decays are negligible in this study. 
The widths of the $\st_1$ and $\sb_1$ 
decays into $H^\pm$ receive large SUSY-QCD corrections for large 
$\tan\b$ in the on-shell renormalization scheme \cite{sq_to_H_QCDcorr}:
the $O(\a_s)$ correction terms are often comparable to or even larger 
than the lowest order ones. Such large corrections make the perturbation 
calculation of the decay widths unreliable. In general this problem shows 
up in calculating rates of processes involving the bottom-Yukawa-coupling 
($h_b$) for large $\tan\b$ in the MSSM \cite{improvedQCDcorr}. In Ref.
\cite{improvedQCDcorr} it is pointed out that this problem can be solved by 
carefully defining the relevant tree-level couplings in terms of appropriate 
running parameters and on-shell squark mixing angles $\t_{\sq}$. 
Following Ref.\cite{improvedQCDcorr}, we calculate the tree-level 
widths of the $\st_1$ and $\sb_1$ decays, Eqs.(\ref{eq:fmodes}) and 
(\ref{eq:bmodes}), by using the corresponding tree-level couplings 
defined in terms of the SUSY-QCD running parameters $m_q(Q)$ and $A_q(Q)$ 
(with the renormalization scale Q taken as the on-shell (pole) mass of the 
decaying squark), and the on-shell squark mixing angles $\t_{\sq}$. 
For the kinematics, i.e. for the phase space factor $\kappa / m_{\sq_1}^3$ 
(for $\sq_1$ decay) the on-shell masses are used. We call the widths thus 
obtained as 'renormalization group (RG) improved tree-level widths'. 
Here we adopt the notations and conventions of \cite{improvedQCDcorr}. 
Our input parameters are all on-shell ones except $A_b$ which is a 
running one, i.e. they are $M_t,M_b,M_{\ti Q}(\st),M_{\ti U},M_{\ti D},
A_t,A_b(Q),\mu,\tan\b,m_A$, and M, with the renormaliztion scale Q at 
the on-shell (OS) mass of the decaying squark $m_{\sq_1 OS}$. 
$M_{t,b}$ are the on-shell (pole) masses of the t,b quarks. $M_{\ti Q}(\sq)$ 
is the on-shell $M_{\ti Q}$ for the $\sq$ sector. Note that $M_{\ti Q}(\st)$ 
is different from $M_{\ti Q}(\sb)$  by finite QCD corrections 
\cite{improvedQCDcorr}. The procedure for obtaining all necessary on-shell and 
$\overline{DR}$-running parameters of quarks and squarks (i.e. on-shell (pole) 
$m_{\sq_{1,2}}$, on-shell $\t_{\sq}$, and SUSY-QCD running 
$(m_q, A_t, M_{\ti Q}, M_{\ti U}, M_{\ti D}, m_{\sq_{1,2}})$) is described 
in detail in \cite{improvedQCDcorr}. We use this procedure. 
For the SM parameters we take $M_t$=175GeV, 
$M_b$=5GeV, $m_Z$=91.2GeV, $\sin^2\t_W=0.23, m_W=m_Z\cos\t_W, 
\a(m_Z)=1/129$, and $\a_s(m_Z)=0.12$ [with the one-loop running 
$\a_s(Q)=12\pi/((33-2n_f)\ln(Q^2/\Lambda_{n_f}^2))$, 
$n_f$ being the number of quark flavors; only for the calculation of 
the SM running quark mass $m_q(Q)_{SM}$ from the two-loop renormalization 
group equations we use the two-loop running $\a_s(Q)$ as in 
\cite{improvedQCDcorr}]. 
In order not to vary too many parameters 
we choose $M_{\ti Q}(\st)$ = $\frac{3}{4} M_{\ti U}$ = $\frac{3}{2} M_{\ti D}$ 
($M_{\ti Q}(\st)$ = $\frac{3}{2} M_{\ti U}$ = $\frac{3}{4} M_{\ti D}$) 
for $\st_1$ ($\sb_1$) decays, and  $A_t=A_b(Q) \equiv A$ 
(where Q=$m_{\sq_1 OS}$ for $\sq_1$ decay) for simplicity. 
Moreover, we fix M=400GeV (i.e. $m_{\sg}$=1065GeV) and $m_A$=150GeV. 
Thus we have $M_{\ti Q}(\st)$, A, $\mu$ and $\tan\b$ as free parameters. 

In the plots we impose the following conditions:
\renewcommand{\labelenumi}{(\roman{enumi})} 
\begin{enumerate}
  \item $m_{\ch_1} > 100$ GeV, $m_{h^0} > $ 110 GeV, 
        $m_{\st_1,\sb_1 OS} > m_{\nt_1}$, 
        $m_{\nt_1} > 90$ GeV, 
  \item $\delta\rho\,(\st\!-\!\sb) < 0.0012$ \cite{deltarho} 
        using the formula of \cite{Drees-Hagiwara} 
        (the constraint from electroweak $\delta\rho$ bound on 
        $\st$ and $\sb$), and
  \item $A_t^2(Q) < 3\,(M_{\ti Q}^2(Q) + M_{\ti U}^2(Q) + m_{H_2}^2)$, and 
        $A_b^2(Q) < 3\,(M_{\ti Q}^2(Q) + M_{\ti D}^2(Q) + m_{H_1}^2)$ with 
        $m_{H_1}^2=(m_A^2+m_Z^2)\sin^2\b-\frac{1}{2}\,m_Z^2$, 
        $m_{H_2}^2=(m_A^2+m_Z^2)\cos^2\b-\frac{1}{2}\,m_Z^2$, and 
        $ Q \sim M_{\ti Q}$ (the approximate necessary condition 
        for the tree-level vacuum stability \cite{stability}).
\end{enumerate}
Conditions (i), along with $m_{\sg} = 1065$ GeV, ensure that we satisfy the 
experimental bounds on $\ti\chi_1^+$, $\ti\chi_1^0$, $h^0$, $\st_1$, $\sb_1$ 
and $\sg$ from LEP2 \cite{LEP} and Tevatron \cite{Tevatron}.  
Note that $m_{\nt_1} > 90$ GeV is imposed in order to evade the experimental 
bounds on $m_{\st_1,\sb_1}$. Conditions (ii) and (iii) constrain the $\st$ and 
$\sb$ mixings significantly. We note that the experimental data for 
the $b \to s \gamma$ decay give rather strong constraints \cite{Tata} 
on the SUSY and Higgs parameters within the minimal supergravity model, 
especially for large $\tan\beta$. However, we do not impose  this 
constraint since it strongly depends on the detailed properties of 
the squarks, including the generation-mixing.

In Fig.1 we plot in the A-$\mu$ plane the contours of the $\st_1$ 
decay branching ratios of the Higgs boson mode $B(\st_1 \to \sb_1 + H^+)$, 
the gauge boson mode $B(\st_1 \to \sb_1 + W^+)$, and the total 
bosonic modes $B(\st_1 \to \sb_1 + (H^+,W^+)) \equiv 
B(\st_1 \to \sb_1 + H^+) + B(\st_1 \to \sb_1 + W^+)$  
at the RG-improved tree-level for $\tan\beta$=30 and 
$M_{\ti Q}(\st)$=600GeV. 
We show also those of $B(\st_1 \to \sb_1 + (H^+,W^+))$ at the naive 
(unimproved) tree-level, where all input parameters are bare ones, i.e. 
they are $m_t$(=175GeV), $m_b$(=5GeV), 
$M_{\ti Q}$ = $\frac{3}{4} M_{\ti U}$ = $\frac{3}{2} M_{\ti D}$ (=600GeV), 
$A_t=A_b \equiv A$, $\mu$, $\tan\b$ (=30), $m_A$(=150GeV), and M(=400GeV) 
(see Eqs.(\ref{eq:f})-(\ref{eq:i})). 
We see that the $\st_1$  decays into bosons are dominant in a large region of 
the A-$\mu$ plane, especially for large $|A|$ and/or $|\mu|$, as we 
expected. Comparing Fig.1.c with Fig.1.d we find that the effect 
of running of the quark and squark parameters 
$(m_q(Q),A_q(Q),M_{\ti Q,\ti U,\ti D}(Q))$ is quite dramatic. 
Note here that the symmetry under $A \to -A$ and/or 
$\mu \to -\mu$, which holds at the naive tree-level, is strongly broken 
at the RG-improved tree-level. This is mainly due to the running 
effect steming from the gluino loops \cite{improvedQCDcorr}.

In Fig.2 we show the contours of the $\sb_1$  decay branching ratios 
$B(\sb_1 \to \st_1 + H^-)$,$B(\sb_1 \to \st_1 + W^-)$, and 
$B(\sb_1 \to \st_1 + (H^-,W^-))$ at the RG-improved tree-level 
for $\tan\b$=8 and 30, and $M_{\ti Q}(\st)$=600GeV. 
The results are very similar to those for the $\st_1$ decays. 
We have found again a dramatic effect of the running of the parameters.

In Fig.3 we show the individual branching ratios of the $\st_1$ and $\sb_1$ 
decays as a function of $\tan\b$ 
for $(A, \mu, M_{\ti Q}(\st))$=(-800, -700, 600)GeV and (800, 800, 600)GeV, 
respectively. One can see that the branching ratios of the $\st_1$  decays 
into bosons increase with increasing $\tan\b$ and become dominant for 
large $\tan\b$ $(\gsim 20)$, while the $\sb_1$ decays into bosons are 
dominant in the entire range of $\tan\b$  shown, as expected. 
As already explained, these $\tan\b$ dependences of the $\st_1$ and $\sb_1$ 
decays come from the increase of $(h_b, a_b m_b)$ with $\tan\b$ and 
the mild dependence of $(h_t, a_t m_t)$ on $\tan\b$, respectively. 

In Fig.4 we show the $M_{\ti Q}(\st)$ dependence of the $\st_1$ and $\sb_1$ 
decay branching ratios for (A(GeV),$\mu$(GeV),$\tan\b$)=(-800, -700, 30) and 
(800, 800, 30), respectively. In these cases we have $(m_{\nt_1},m_{\chp_1},
m_{\sg})$=(198, 393, 1065)GeV and (198, 394, 1065)GeV, respectively. 
One can see that the bosonic modes dominate the $\st_1$ and $\sb_1$ decays 
in a wide range of $M_{\ti Q}(\st)$. For the $\sb_1$ decays we have obtained 
a similar result for (A(GeV), $\mu$(GeV), $\tan\b$)=(800, -800, 8). 
(Note that the decay into a gluino becomes dominant above its threshold.) 
One can also see that the branching ratios of the bosonic decays decrease 
with increasing $M_{\ti Q}(\st)$ which is roughly equal to the mass of 
the decaying squark $\sq_1$. This comes from the fact that in the large 
$M_{\ti Q}(\st)(\sim m_{\sq_1})$ limit the decay widths of the bosonic 
and fermionic modes are proportional to $m_{\sq_1}^{-1}$ and $m_{\sq_1}$, 
respectively.

We find that the dominance of the bosonic modes is fairly insensitive 
to the choice of the values of $m_A$, M, and the ratio $A_b(Q)/A_t$. 
The decays into $H^\pm$ are kinematically suppressed for large $m_A$. 
However, the remaining gauge boson mode can still be dominant. 
For example, for $M_{\ti Q}(\st)$=600GeV, A=$\mu$=800GeV, and $\tan\b$=30, 
we have 
$(m_{\st_1 OS}, m_{\sb_1 OS}, m_{\nt_1}, m_{\chp_1})$=(347, 602, 198, 394)GeV 
with $B(\sb_1 \to \st_1 \, H^-)$=(48, 37, 13, 0)\% and 
$B(\sb_1 \to \st_1 \, W^-)$=(43, 51, 71, 82)\% 
for $m_A$=(120, 200, 240, $>$250)GeV. 
We have also checked that our results do not change 
significantly if we take smaller values of M, as long as decays into a gluino 
are kinematically forbidden. As for the sensitivity of our results to the 
ratios $M_{\ti U,\ti D}/M_{\ti Q}(\st)$, we have found that for small 
$\tan\b$ and large $|A|$ and $|\mu|$ the bosonic modes can dominate the 
$\sb_1$ decay even in case $M_{\ti Q}(\st)=M_{\ti U}=M_{\ti D}$; e.g. for 
$\tan\b=8$ and $M_{\ti Q}(\st)=M_{\ti U}=M_{\ti D}=600$GeV, we have obtained 
results similar to Fig.2.b and Fig.2.c. For example we have 
$B(\sb_1 \to \st_1 + W^-)=60\%$, $B(\sb_1 \to \st_1 + (H^-,W^-))=90\%$ and 
$(m_{\st_1 OS}, m_{\st_2 OS}, m_{\sb_1 OS}, m_{\sb_2 OS}, m_{H^+})=
(388, 792, 570, 638, 166)$GeV for A=1200GeV, $\mu$=-1300GeV, $\tan\b=8$, 
and $M_{\ti Q}(\st)=M_{\ti U}=M_{\ti D}=600$GeV. For these parameters the 
$\st_L$-$\st_R$ mixing term is large and the $\sb_L$-$\sb_R$ mixing term is 
small. This leads to a large mass-splitting of $\st_1$-$\st_2$ and a small 
one of $\sb_1$-$\sb_2$, which results in sufficient phase spaces for 
the bosonic decays of $\sb_1$. 

In a complete analysis one would have to calculate the full SUSY-QCD one-loop 
corrections to the widths of the $\st_1$ and $\sb_1$ decays. 
However, we expect that these corrections will not invalidate 
the dominance of the bosonic modes in the $\st_1$ and $\sb_1$ decays in a 
significant portion of the MSSM parameter space as shown in our analysis. The 
calculation of the full corrections is beyond the scope of the present paper. 

Now we discuss the signatures of the $\st_1$ and $\sb_1$ decays.
We compare the signals of the decays into bosons (Eq.~(\ref{eq:bmodes})) 
with those of the decays into fermions (Eq.~(\ref{eq:fmodes})). 
In principle, the final states of the bosonic decays 
can also be generated from fermionic decays. 
For example, the final particles of the decay chain 
\beq \label{chaina}
\st_1 \to \sb_1 + (H^+ \; {\rm or} \; W^+) \to 
(b \nt_1)+(q \bar q^\prime) 
\eeq
are the same as those of 
\beq \label{chainb} 
\st_1 \to b + \chp_{1,2} \to b + ((H^+ \; {\rm or} \; W^+) + \nt_1) \to 
b + (q \bar q^\prime \nt_1).
\eeq
Nevertheless, the decay distributions of the two processes 
(\ref{chaina}) and (\ref{chainb})  are in general different from 
each other due to the different intermediate states. 
For example, the b in the chain (\ref{chaina}) tends to be softer 
than the b in (\ref{chainb}). A similar argument holds for the 
quark pairs $q \bar q^\prime$ in the decay chains. 
Moreover, the distribution of the missing energy-momentum carried 
by $\nt_1$ could be significantly different in (\ref{chaina}) and 
(\ref{chainb}) since it is emitted from a different sparticle. 
Hence the possible dominance of the bosonic decays over the conventional 
fermionic decays could have an important impact on the search for the 
$\st_1$ and $\sb_1$, and on the measurement of the MSSM parameters. 
Therefore, the effects of the bosonic decays should be included in 
the Monte Carlo studies of the $\st_1$ and $\sb_1$ decays.


In conclusion,   
we have shown that the $\st_1$ and $\sb_1$ decays into Higgs or 
gauge bosons, such as $\st_1$ $\to$ $\sb_1$ + $(H^+$~or~$W^+)$, 
can be dominant in a fairly wide MSSM parameter region 
with large mass difference  between $\st_1$ and $\sb_1$, large $|A_{t,b}|$ 
and/or $|\mu|$, and large $m_{\sg}$ (and large $\tan\b$ for the $\st_1$ decay) 
due to the large Yukawa couplings and mixings of $\st$ and $\sb$. 
Compared to the fermionic decays, such as 
$\st_1 \to b + \chp_{1,2}$, these bosonic decays can have significantly 
different decay distributions. We have also shown that the effect of 
the SUSY-QCD running of the quark and squark parameters on the $\st_1$ and 
$\sb_1$ decays is quite dramatic. 
These could have an important impact 
on the searches for $\st_1$ and $\sb_1$ and 
on the determination of the MSSM parameters at future colliders.

\section*{Acknowledgements}
We are very grateful to S. Kraml and H. Eberl for valuable discussions and 
correspondence and for checking some of our numerical results. 
One of the authors (K.H.) would like to thank H. E. Haber, J. L. Hewett, 
W. Majerotto and S. Pokorski for useful discussions and suggestions. He also 
thanks the Theory Groups of Fermilab and SLAC for their hospitality 
during the course of this article. The work of A.B. was supported by the 
``Fonds zur F\"orderung der wissenschaftlichen Forschung of Austria'', 
project no. P13139--PHY, and by the EU contract HPRN-CT-2000-00149.


\newpage



\begin{flushleft}
{\Large \bf Figure Captions} \\
\end{flushleft}

\noi
{\bf Figure 1}: 
Contours of branching ratios of $\st_1$ decays at the RG-improved 
tree-level in the A--$\mu$ plane for $\tan\b=30$, and 
$M_{\ti Q}(\st)$ = $\frac{3}{4} M_{\ti U}$ = $\frac{3}{2} M_{\ti D}$ = 600GeV; 
(a)$B(\st_1 \to \sb_1 + H^+)$, (b)$B(\st_1 \to \sb_1 + W^+)$, and 
(c)$B(\st_1 \to \sb_1 + (H^+,W^+))$. The regions outside of the dashed loops 
are excluded by the kinematics and/or the conditions (i) to (iii) given in 
the text. Contours of the corresponding branching ratio 
$B(\st_1 \to \sb_1 + (H^+,W^+))$ at the naive tree-level are shown in Fig.d. 

\noi
{\bf Figure 2}:
Contours of branching ratios of $\sb_1$ decays at the RG-improved 
tree-level in the A--$\mu$ plane for $\tan\b$= 8 (a-c) and 30 (d-f), and 
$M_{\ti Q}(\st)$ = $\frac{3}{2} M_{\ti U}$ = $\frac{3}{4} M_{\ti D}$ = 600GeV; 
(a,d)$B(\sb_1 \to \st_1 + H^-)$, (b,e)$B(\sb_1 \to \st_1 + W^-)$, and 
(c,f)$B(\sb_1 \to \st_1 + (H^-,W^-))$. The regions outside of the dashed loops 
are excluded by the kinematics and/or the conditions (i) to (iii) given in 
the text. 

\noi
{\bf Figure 3}:
$\tan\b$ dependence of $\st_1$ (a) and $\sb_1$ (b) decay branching ratios 
for $(A, \mu, M_{\ti Q}(\st))$ = (-800, -700, 600)GeV and (800, 800, 600)GeV,
respectively. "$\sb_1 + H^+/W^+$" and "$\st_1 + H^-/W^-$" refer to the sum 
of the Higgs and gauge boson modes. The gray areas are excluded by the 
conditions (i) to (iii) given in the text. 

\noi
{\bf Figure 4}:
$M_{\ti Q}(\st)$ dependence of $\st_1$ (a) and $\sb_1$ (b) decay branching 
ratios for (A(GeV), $\mu$(GeV), $\tan\b$) = (-800, -700, 30) and (800, 800, 30),
respectively. "$\sb_1 + H^+/W^+$" and "$\st_1 + H^-/W^-$" refer to the sum 
of the Higgs and gauge boson modes. The gray areas are excluded by the 
conditions (i) to (iii) given in the text. 

\newpage
%
%
%
\begin{figure}[!htb]
\begin{center}
\vspace{20mm}
\scalebox{0.6}[0.6]{\includegraphics{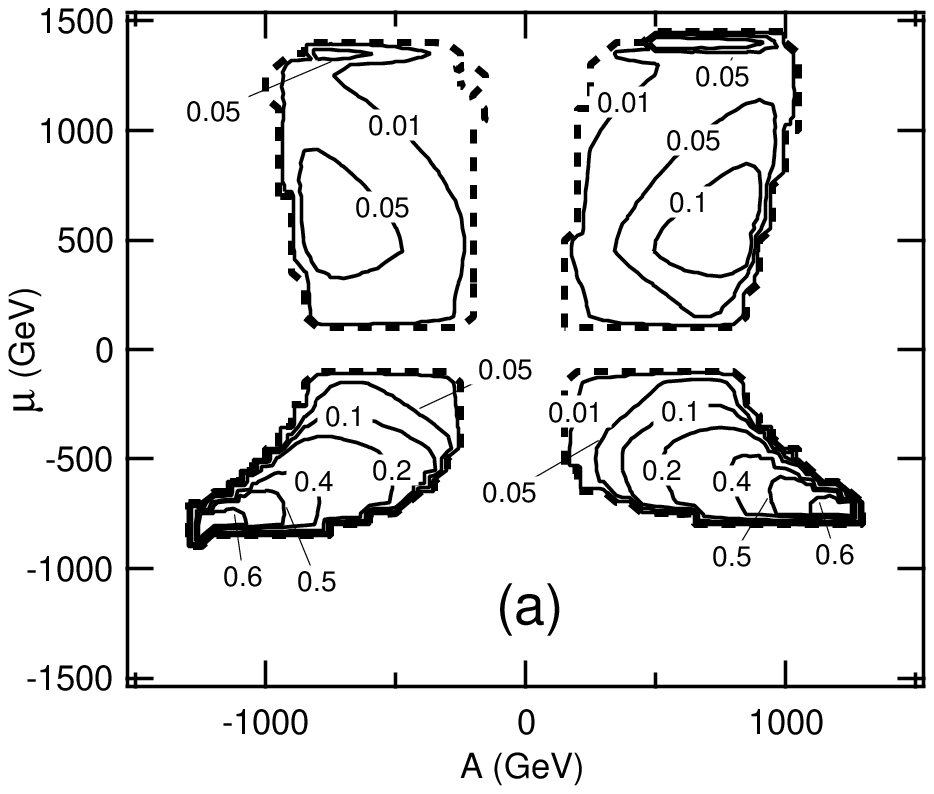}} 
\scalebox{0.6}[0.6]{\includegraphics{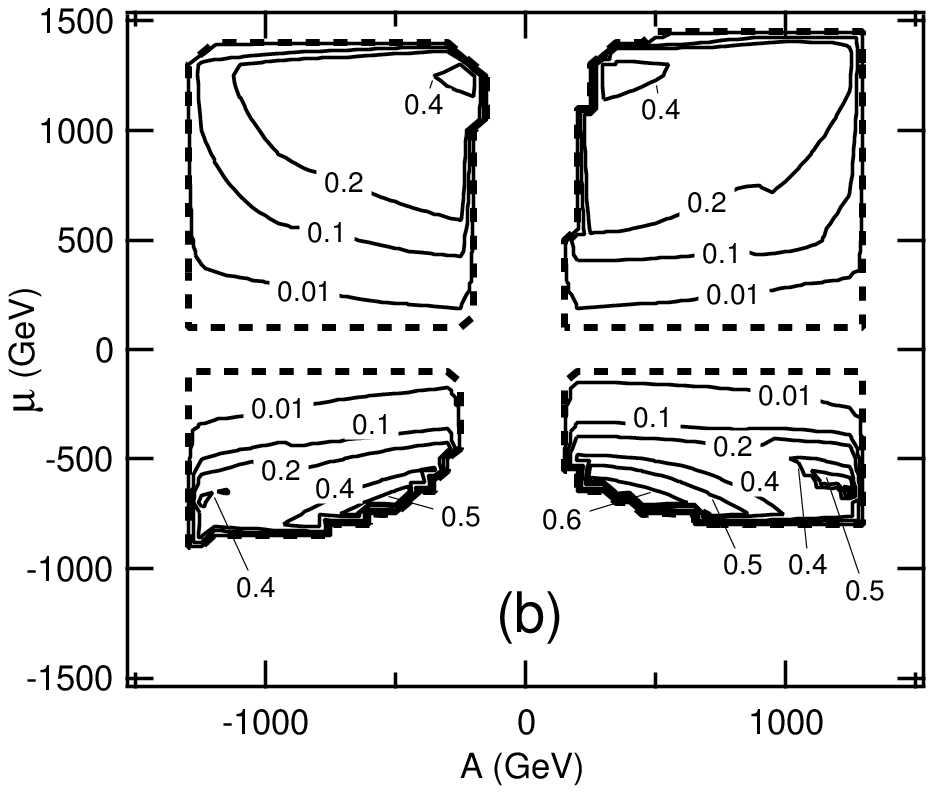}} \\
\scalebox{0.6}[0.6]{\includegraphics{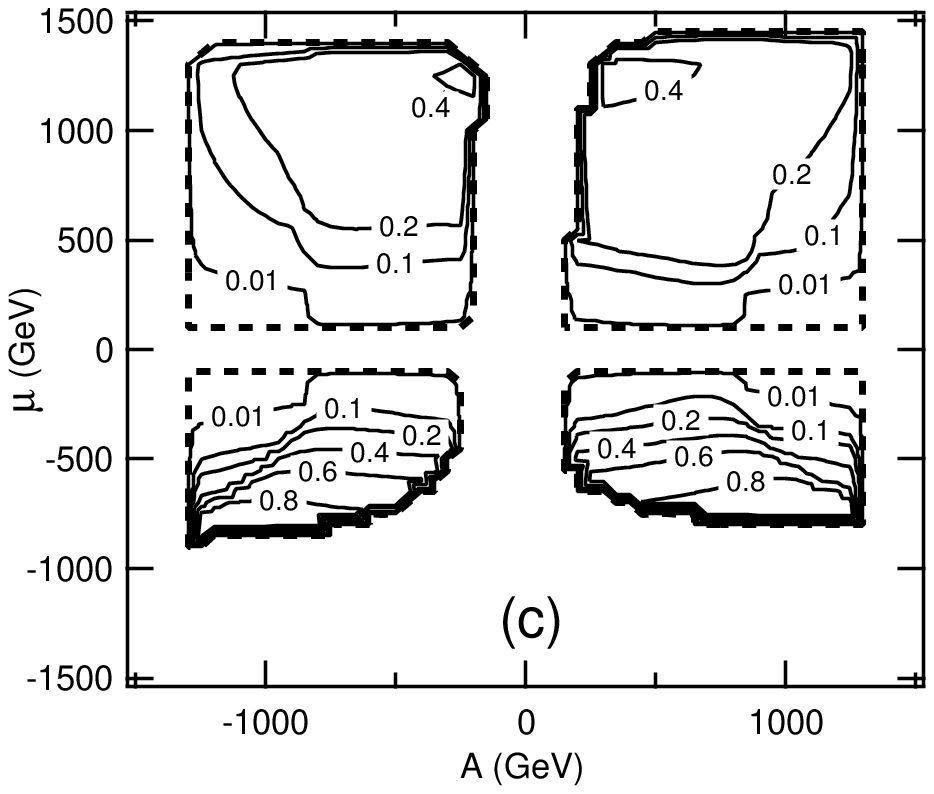}} 
\scalebox{0.6}[0.6]{\includegraphics{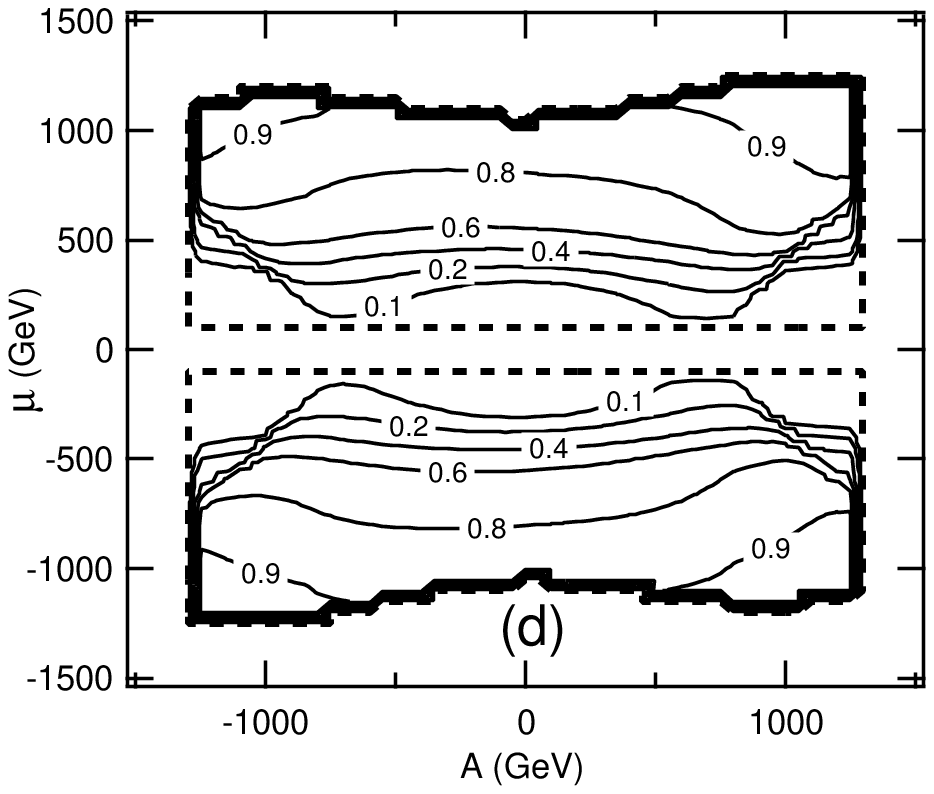}} \\
\vspace{5mm}
{\LARGE \bf Fig.1}
\end{center}
\end{figure}
%

\newpage
%
%
\begin{figure}[!htb]
\begin{center}
\scalebox{0.6}[0.6]{\includegraphics{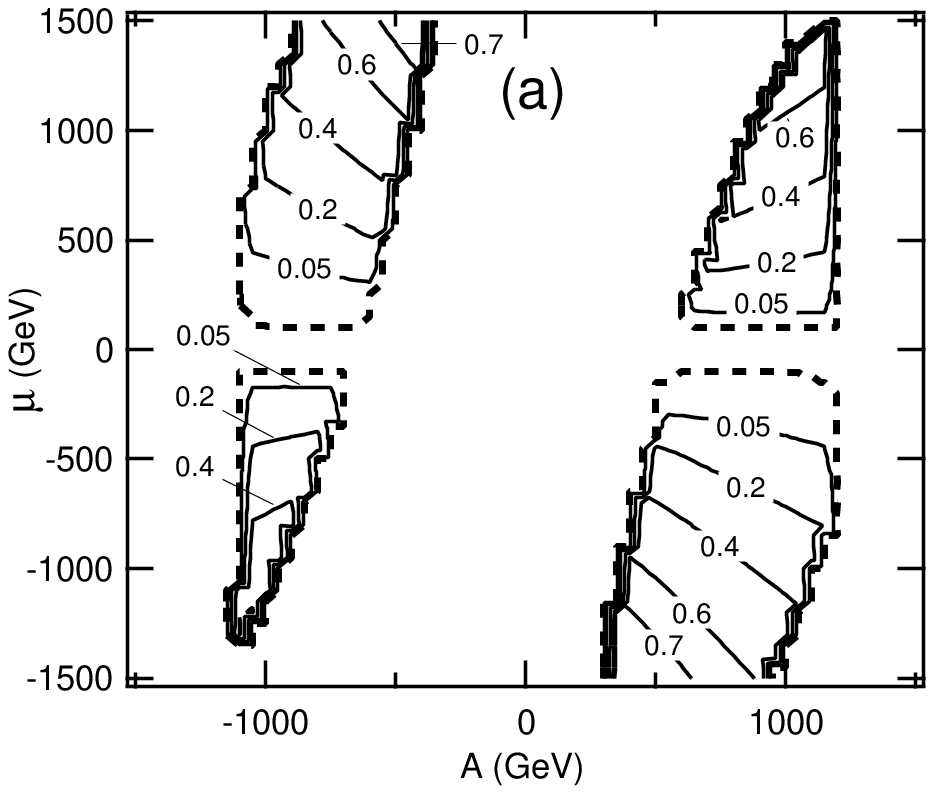}} 
\scalebox{0.6}[0.6]{\includegraphics{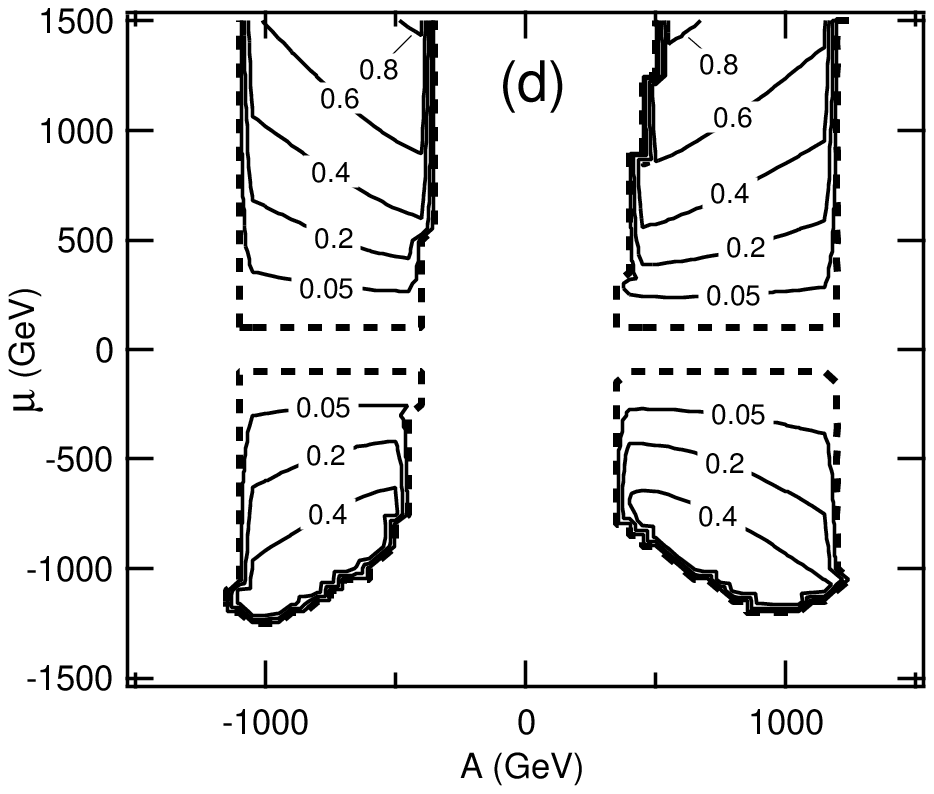}} \\
\scalebox{0.6}[0.6]{\includegraphics{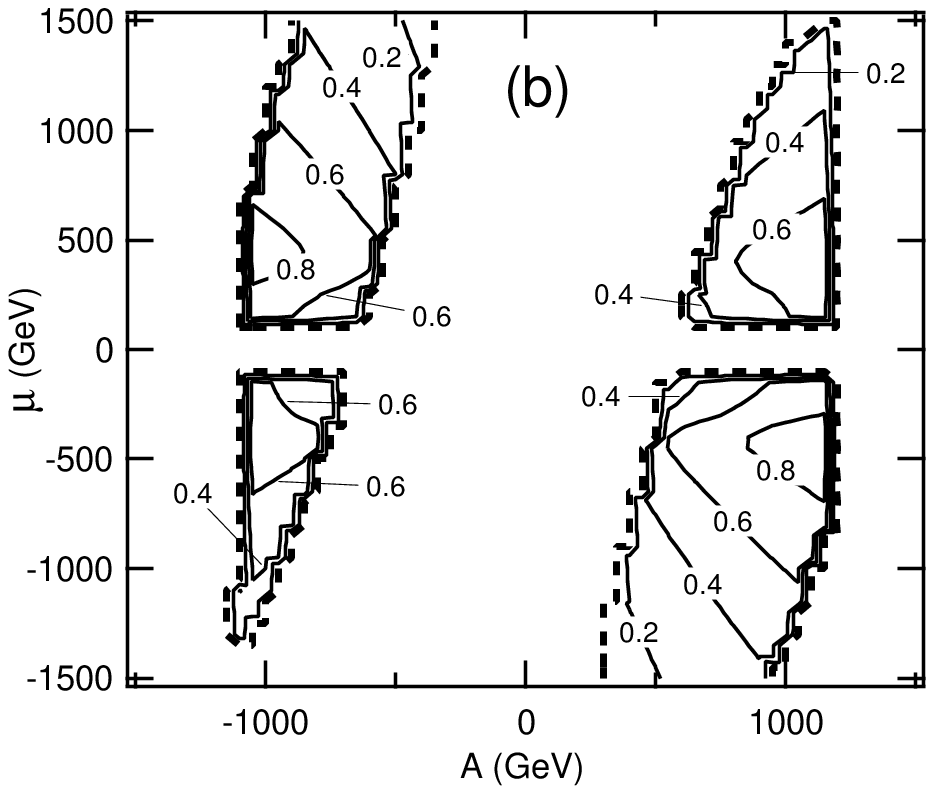}} 
\scalebox{0.6}[0.6]{\includegraphics{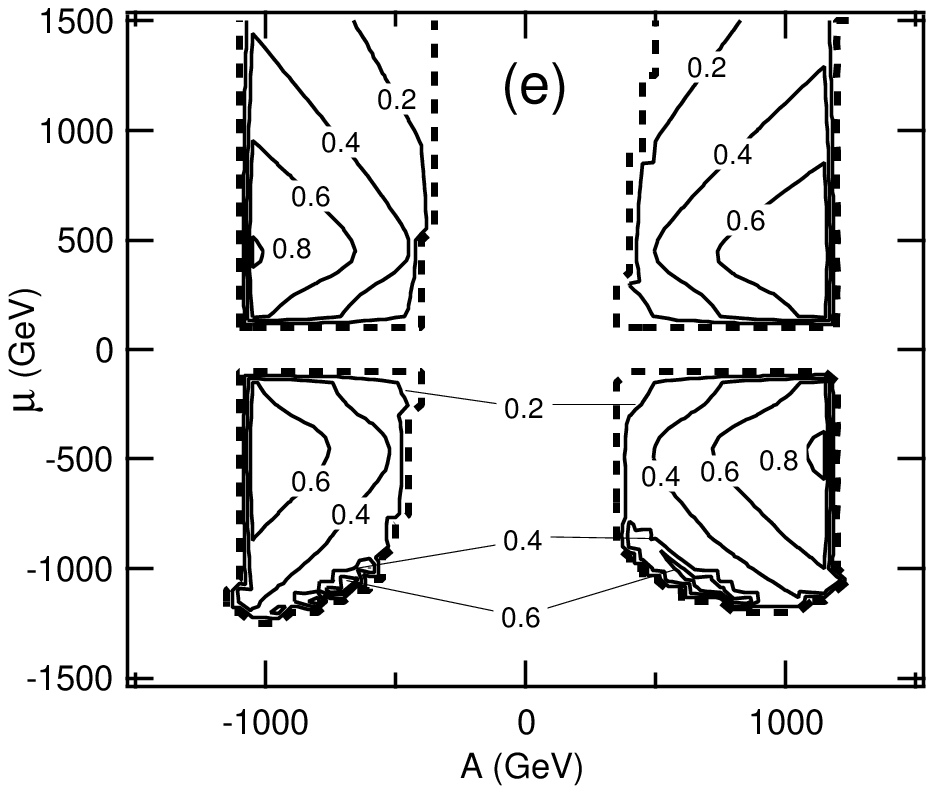}} \\
\vspace{2mm}
\scalebox{0.6}[0.6]{\includegraphics{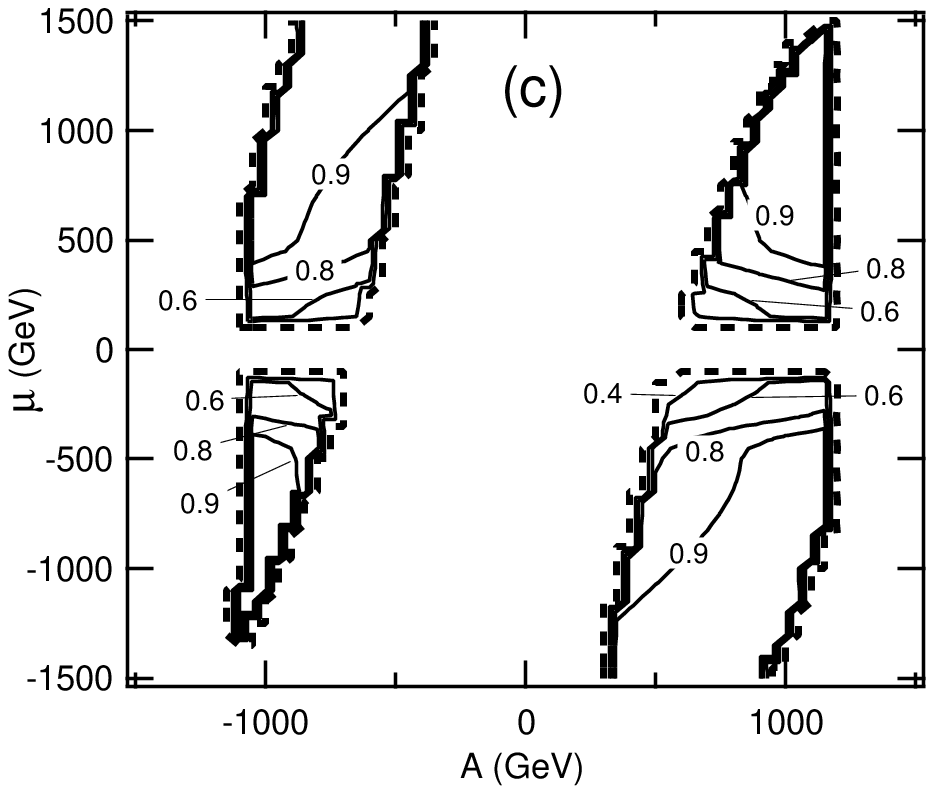}} 
\scalebox{0.6}[0.6]{\includegraphics{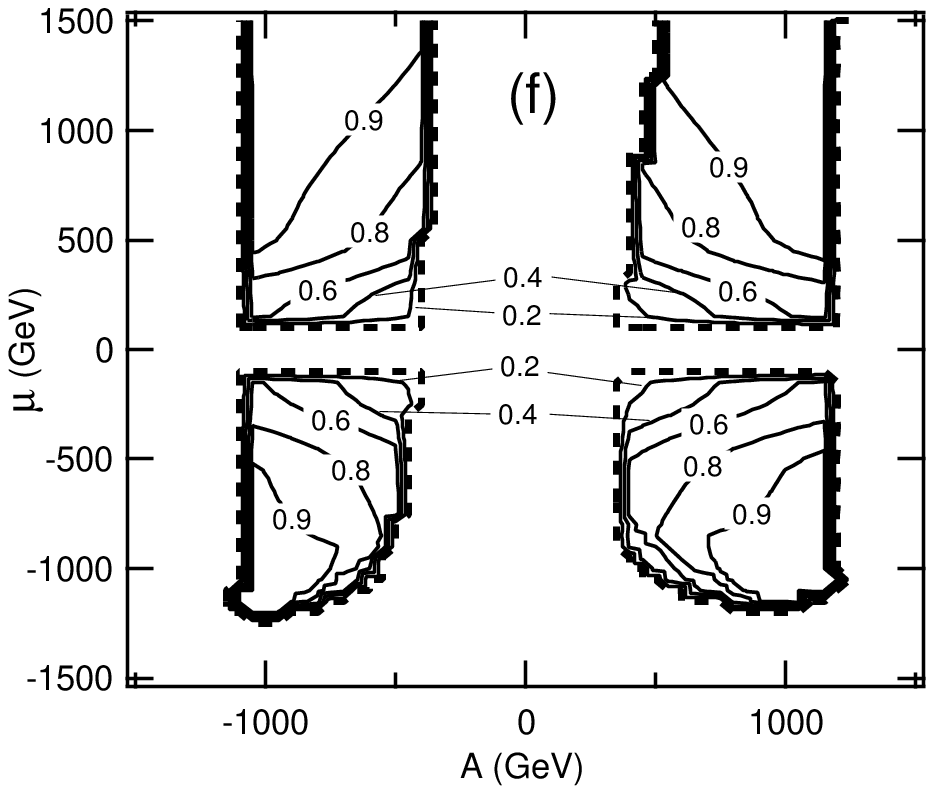}} \\
\vspace{5mm}
{\LARGE \bf Fig.2}
\end{center}
\end{figure}

\newpage
%
%
\begin{figure}[!htb]
\begin{center}
\vspace{5mm}
\scalebox{1.0}[1.0]{\includegraphics{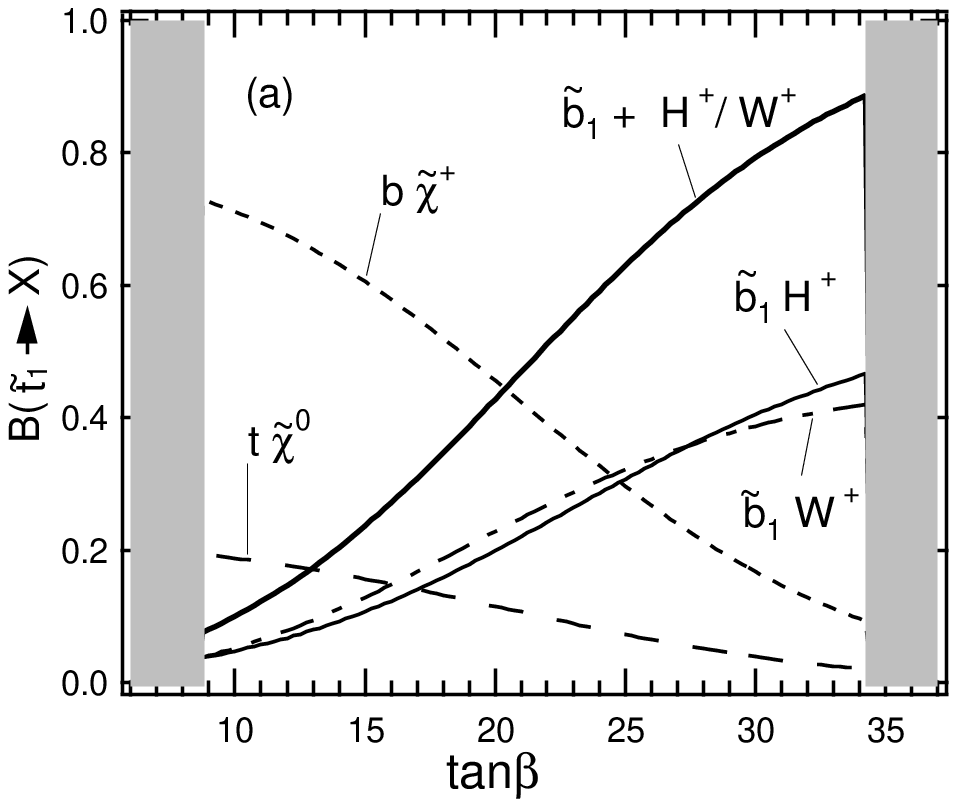}} \\
\vspace{5mm}
\scalebox{1.0}[1.0]{\includegraphics{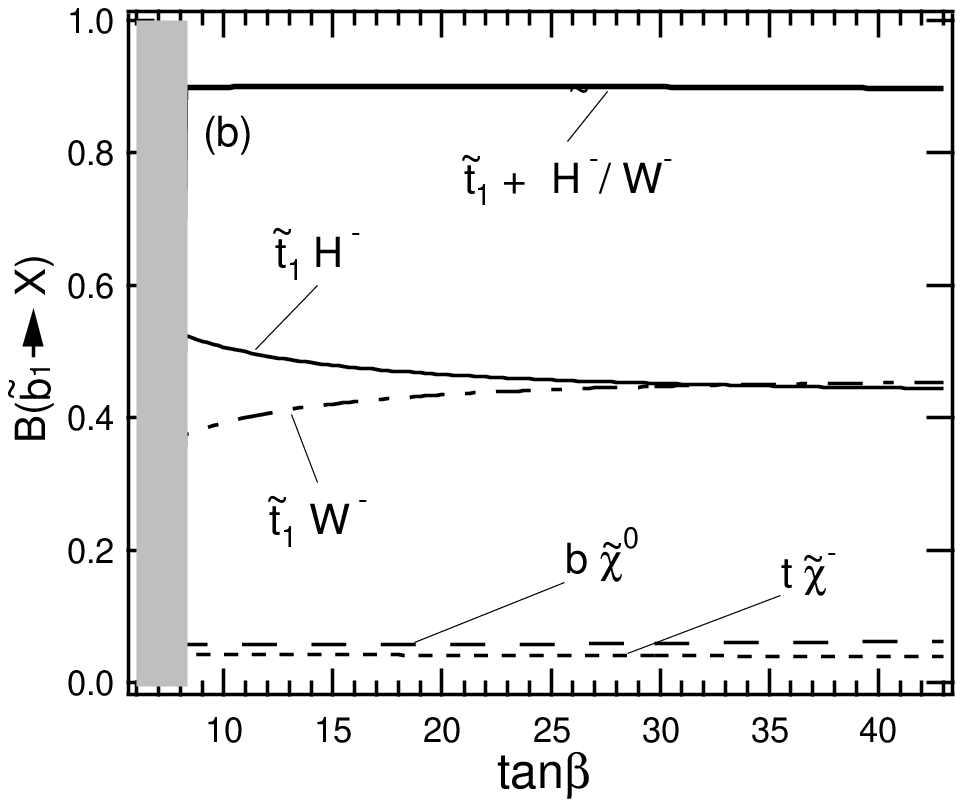}} \\
\vspace{10mm}
{\LARGE \bf Fig.3}
\end{center}
\end{figure}

\newpage
%
%
\begin{figure}[!htb]
\begin{center}
\vspace{5mm}
\scalebox{1.0}[1.0]{\includegraphics{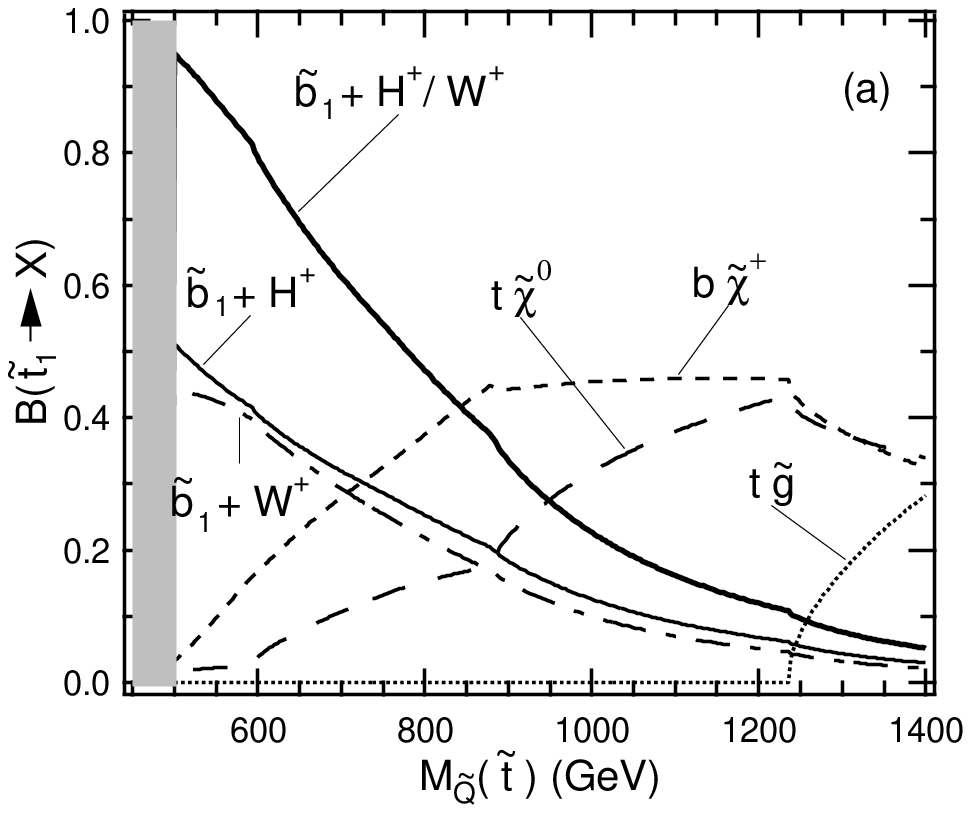}} \\
\vspace{5mm}
\scalebox{1.0}[1.0]{\includegraphics{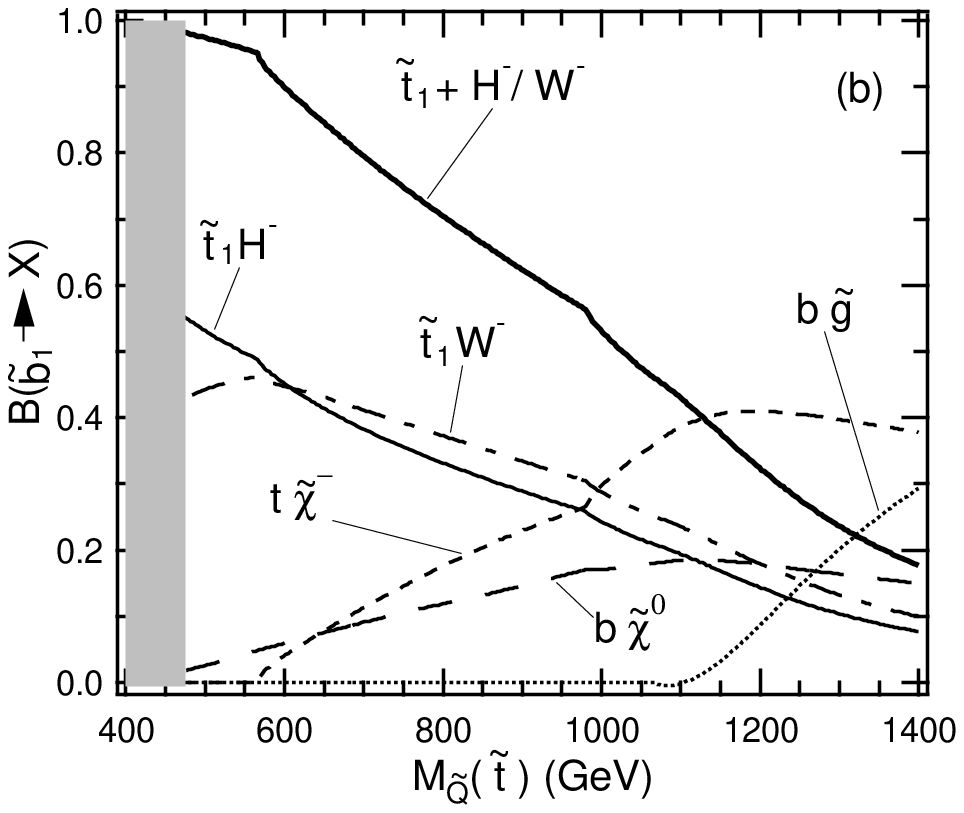}} \\
\vspace{10mm}
{\LARGE \bf Fig.4}
\end{center}
\end{figure}
%


\end{document}